\documentclass[a4paper,20pt]{article}
    \usepackage[T1]{fontenc}
    \usepackage{epsfig}
    \usepackage{graphicx}
    \usepackage{amsmath}
    \usepackage{amsfonts}
    \renewcommand{\abstract}{}
\begin{document}
\makeatletter
\renewcommand{\@oddhead}{\textit{YSC'14 Proceedings of Contributed Papers} \hfil \textit{N.M.Kostogryz}}
\renewcommand{\@evenfoot}{\hfil \thepage \hfil}
\renewcommand{\@oddfoot}{\hfil \thepage \hfil}
\fontsize{11}{11} \selectfont

\title{Study of the Reasons for the Geometric Albedo Variations of Uranus}
\author{\textsl{N.M.~Kostogryz}}
\date{}
\maketitle
\begin{center} {Main Astronomical Observatory of NAS of Ukraine,\\
27, Zabolotnoho str., Kyiv, Ukraine, 03680 \\kosn@mao.kiev.ua\\ }
\end{center}

\begin{abstract}
The method of the optical parameter estimations of the
nonisothermal giant planet atmospheres by using intensity data of
Raman scattering features was used. We applied this method to the
observational data of Uranus' geometric albedo spectra from 1981,
1993 and 1995 and obtained the spectral values of the optical
depth, namely, $\tau_a/\tau_R$ and $\tau_\kappa/\tau_S$ (where
$\tau_a$, $\tau_R$ are aerosol and gas components,
$\tau_S=\tau_a+\tau_R$ and $\tau_\kappa$ is absorbtion component
of the effective optical depth of the intensity of diffuse -
reflected irradiation forming). We showed that these ratios are
different for the three years. The conclusion is that this effect
can be due to the horizontal inhomogeneity of aerosol component of
optical depth over the Uranus' disk.
\end{abstract}

\section*{Introduction}

\indent \indent Among giant-planets Uranus is unique from several aspects.
Firstly, its inner energy source is very weak, so Uranus might be
expected to have small amount of the atmospheric turbulence.
Secondly, uranian rotational pole lies almost in its orbital
plane. This anomalous obliquity allows us to obtain images of
different uranian latitudes and causes radical variations in
insolation between seasons, in comparison to the other planets.

In recent years, the observational data of the intensity of the
Raman scattering features in the giant planet spectra are
suggested to be used in determining the relative contribution of
the aerosol component of the atmosphere. This way, we can
determine the values of aerosol to gas ratio of the optical depth
components $\tau_a/\tau_R$ and absorbing to scattering ratio
$\tau_\kappa/\tau_R$ \cite{Dementiev1, Morozhenko1}. Dementiev
\cite{Dementiev1} showed that in Uranian atmosphere the aerosol
abundance was much larger in 1981-1983 than in 1961-1973. In these
papers the model of atmosphere was taken to be isothermal, while
the real giant planet atmospheres have complex temperature
profiles \cite{Lindal}. The method of the real temperature profile
accounting used in computing the Raman scattering effects was
developed by Morozhenko and Kostogryz \cite{Morozhenko4}. They
estimated the distortion degree of the atmospheric parameters such
as $\tau_a/\tau_R$ and $\tau_\kappa/\tau_R$. In the paper
\cite{Kostogryz} the real values of these parameters were
estimated using observational geometric albedo spectra
\cite{Karkoschka1}. We should note that Sromovsky
\cite{Sromovsky1} is also taking into account the real temperature
profile. He compared the calculated geometric albedo values with
the observational ones.

In this paper we determine the values of the optical parameters
such as $\tau_a/\tau_R$ and $\tau_\kappa/\tau_R$ using
observational data in different years \cite{Karkoschka1, Karkoschka2, Neff}.

Section 2 describes the observational data for Uranus. Section 3
contains the review of the atmospheric model. Section 4 is devoted to
the method of computation and sections 5 and 6 describe the results of
the computation and some conclusions of this work.

\section*{Observational data}

\indent \indent To obtain the optical parameters of Uranus' atmosphere, we used
geometric albedo spectra observed by Neff et al. \cite{Neff} in
1981 and by Karkoschka \cite{Karkoschka1, Karkoschka2} in 1993 and
1995.

Neff et al. \cite{Neff} derived absolute measurements of the
geometric albedo spectra of Uranus from 350 to 1050 nm at the
resolution of about 0.7 nm. These observations were made with
McDonald Observatory 2.1-m telescope and the ES-2 spectrograph on
17-18 May 1981 UT. The Uranus' radius is taken to be 25700 km,
when the geometric albedo spectra were calculated.

Full-disk albedo of Uranus derived by Karkoschka from
observations at the European Southern Observatory in July 1993
\cite{Karkoschka1} and in July 1995 \cite{Karkoschka2}. The
spectra extend from 300- to 1000-nm wavelength at 1-nm resolution.
In 1995 the spectra extend from 300 to 1050 nm wavelength. The
spectral resolution was 0.4 nm between 520 and 995 nm, and 1 nm
elsewhere. Planetary radius was taken to be 25450 km.

The geometric albedo spectra of Uranus in spectral region from 390
to 845 nm at 2-nm resolution derived by Dementiev
\cite{Dementiev2} was not considered because of the low resolution
and large errors (near 10$\%$) in violet region of the spectra.

For our investigations in all observational data we've just used
the spectral region from 350 to 430 nm where the methane band are
very weak or almost absent.

The differences in geometric albedo spectra are near $10 \%$ for
observational data of Karkoschka \cite{Karkoschka1, Karkoschka2}
and Neff et al.\cite{Neff}. The reason for these changes could be
either real variations in planetary atmosphere or errors made in
the data reduction (for example, for different planetary radii or
comparison stars). As we cannot correct errors for using different
comparison stars, we take that errors arise due to the different
planetary radii. We assumed that visible radius of Uranus was not
changed in comparably short time period and reduce all data to
$R_0 = 25450$ km.

But the corrections for the different radii have small influence
on the geometric albedo time variations. So, most likely, these
distinctions are due to the real variations of the physical
parameters of Uranus' atmosphere. Later on we try to find the
reasons for these differences.

\section*{Model of atmosphere}

\indent \indent  As in previous papers, we used the model of the homogeneous semi-infinite gas-aerosol layer with relative concentration of hydrogen (85$\%$) and helium (15$\%$). We also consider  nonisothermal atmosphere of Uranus using experimental temperature profiles \cite{Lindal}.

Raman scattering is considered for four major hydrogen  transitions: the rotational transitions S(0), S(1) and O(2) and the vibrational one $Q_1(1)$, which produce significant "Raman ghosts". Raman shifts and cross sections at 400 nm of several transitions of molecules of interest in planetary atmospheres were taken from Cochran and Trafton \cite{Cochran}.

At the temperatures of planetary atmospheres, most of the molecules are in the lower rotational levels of the ground vibrational state. Therefore, the Stokes component of the Raman scattering will dominate. The molecule will absorb energy and the photon will emerge at lower frequency, i.e. longer wavelength.

The values of the effective pressure on the intensity where the diffuse reflected irradiation is forming were taken from Morozhenko \cite{Morozhenko3}

\section*{Method of computation}

\indent \indent To take into account the Raman scattering effects, Pollack's equation for the single-scattering albedo was used \cite{Pollack}. Pollack considered only the $Q_1$ vibrational transition, which alone would be unable to generate the ghost features arising from the rotational transitions. Sromovsky \cite{Sromovsky1} compared different models with Raman scattering taken into account. He said that the problem Pollack's original formulation was that it resulted in $\omega > 1$ at some wavelengths, as noted by Courtin \cite{Courtin}, which leads to instability in the solution of the radiation transfer equation. He propose his own approximation for
molecular single scattering albedo. We also propose more precise expression for the single-scattering albedo, and the form given in Eq.{\ref{omega}} taking into account  rotational $(S(0), O(2), S(1))$ and vibrational $(Q_1(1))$ transitions.

\begin{equation}
\omega=\frac{\tau_a/\tau_R +
D}{1+\tau_a/\tau_R+\tau_\kappa/\tau_R} \label{omega}
\end{equation}

\begin{equation}
D=1+0.85*[((N_0\tau_{S(0)}+N_2\tau_{O(2)})f_{{\lambda}_1}+N_1\tau_{S(1)}f_{{\lambda}_2}+\tau_{Q_1(1)}f_{{\lambda}_3})/f_{{\lambda}_0}\tau_R]-A
\label{parD}
\end{equation}

\begin{equation}
A=0.85*(N_0\tau_{S(0)}+N_2\tau_{O(2)}+N_1\tau_{S(1)}+\tau_{Q_1(1)})/\tau_R
\end{equation}
where $f_{{\lambda}_1}, f_{{\lambda}_2}, f_{{\lambda}_3}$ are values of the energy in the Solar spectrum at
the wavelength from which rotation $(S(0), O(2), S(1))$ and vibration $(Q_1(1))$ Stokes transitions of the Raman scattering carries the sun's photon on wavelength $\lambda_0$ accordingly; $\tau_{S(0)}, \tau_{O(2)}, \tau_{S(1)} $ and $\tau_{Q_1(1)}$ are the optical depths of the Raman scattering of the corresponding
transitions, $N_0, N_1, N_2$ - the amount of hydrogen molecule in ortho- and para- state.

Firstly, we estimated the spectral values of the single-scattering  albedo, comparing the observational geometric
albedo spectra \cite{Karkoschka1, Karkoschka2, Neff} with the theoretical one, computed for the homogeneous semi-infinite layer  (Ovsak \cite{Morozhenko2}). We used the Rayleigh phase function. The values of D depend on the effective level of diffuse - reflected irradiation forming. Also we used the approximate values of the effective pressure of this irradiation forming for Uranus' disk in 376-439 nm wavelengths. Afterwords, using the following expression
\begin{equation}
\frac{1}{\omega}=\frac{1+a}{a+D}+\frac{b}{a+D}-1=\left(\frac{\tau_\kappa}{\tau_S}\right)'
\end{equation}
we found the values of $(\frac{\tau_\kappa}{\tau_S})'$ which were anomalous due to the presence of the Raman scattering. Here $a=(\frac{\tau_a}{\tau_R}), b=(\frac{\tau_\kappa}{\tau_R})$. Obviously, b will be changed with the wavelength changing in the case of the absence of the Raman scattering effects (except for the errors of the observational data). The real values of the ratio $(\frac{\tau_\kappa}{\tau_S})$ are \begin{equation}
\frac{\tau_\kappa}{\tau_S}=\frac{[(\frac{\tau_\kappa}{\tau_S})'*a(\lambda)+\frac{D}{\omega_\lambda}-1]}{1+a(\lambda)}
\end{equation}

We supposed that $a_\lambda$ weakly depends on the wavelength in a very narrow spectral region $(\Delta\lambda = 10 nm)$. Thus in the first approximation we neglected this dependence. Then the values of $a$ were varied to obtain the minimum dispersion of $(\frac{\tau_\kappa}{\tau_S})_\lambda$ in each spectral region.

\section*{Results and discussions}

\indent \indent We analyzed the geometric albedo spectra of Uranus in different years in the spectral region 350 - 430 nm and obtained values of $a_\lambda$ and $(\frac{\tau_\kappa}{\tau_S})_\lambda$ (Fig.1). It is necessary to note here that the uncertainties of the geometric albedo determination have effect only on $(\frac{\tau_\kappa}{\tau_S})_\lambda$ determination.
\begin{figure}[h]
\includegraphics{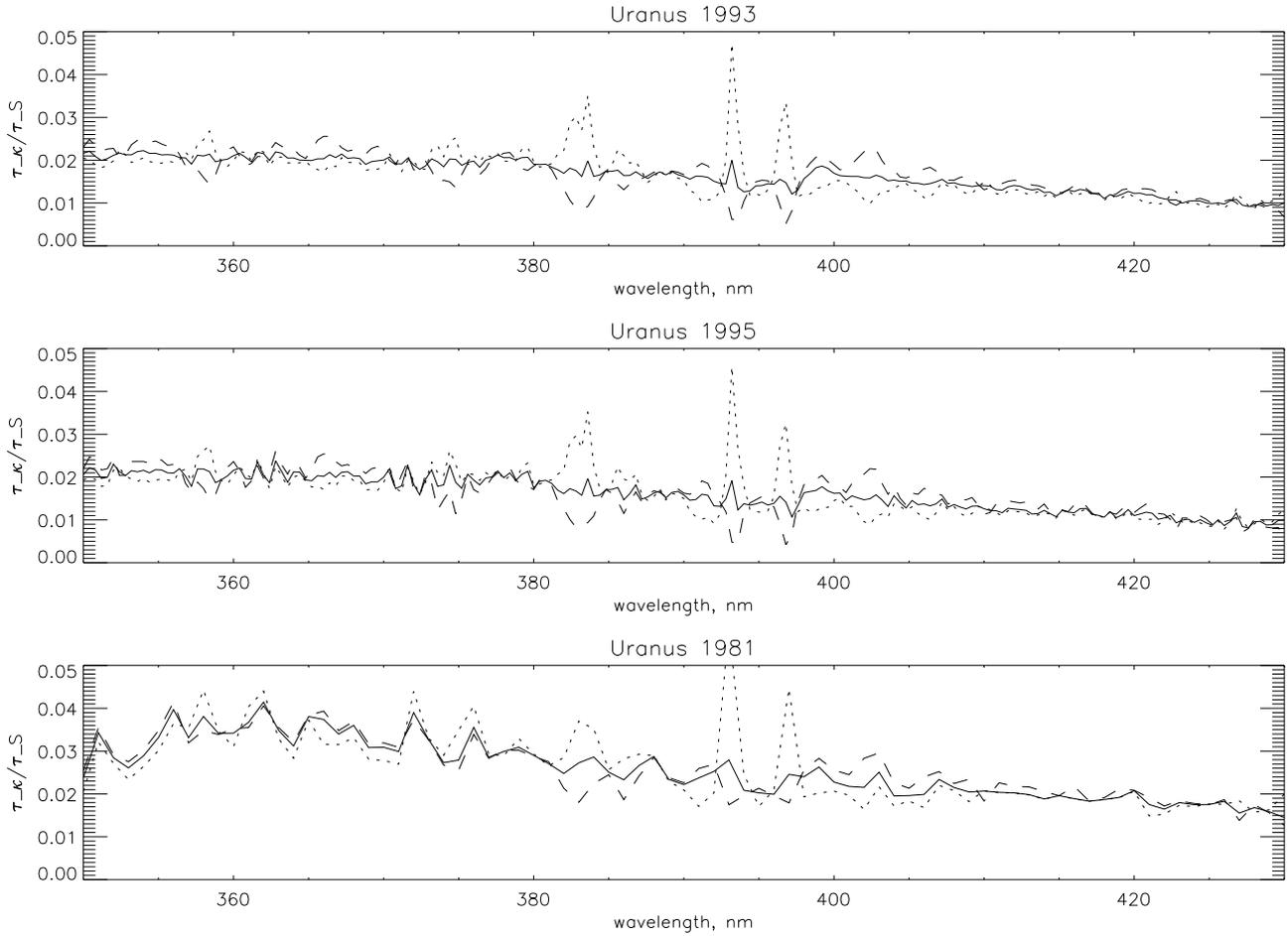} \caption{Spectral dependences of ($\tau_\kappa/\tau_S$) for Uranus in different years.
Observational data marked with dotted line; short dotted line are for the gas atmosphere ($a=0$); and solid line is for the real $(\frac{\tau_\kappa}{\tau_S})_\lambda$} without the Raman scattering influence.
\end{figure}

We can see that in the spectral region where the intensities of the Fraunhofer lines are strong, the values of $a$ are much greater than in the region without those lines. Unfortunately, we are still not able to assume the reason for this phenomena. The possibility that it occurs due to the not fully correct calculations for the Raman scattering effects in our method cannot be excluded.

The observational correlation between $a$ and $(\frac{\tau_\kappa}{\tau_S})$ (Table 1) confirms the supposition
that the temporal variations of the geometric albedo spectra are due to the amount of absorbing aerosol which changes in the planetary atmosphere.
\begin{table}
\caption{The values of $a$ in spectral region from 390 to 399 nm in different years; the values of $(\frac{\tau_\kappa}{\tau_S})$ for 393-nm wavelength and also the ratio of the value of the
geometric albedo for the Fraunhofer line KCaII to its ghost.}
\begin{center}
\begin{tabular}{|c|c|c|c|}
\hline Observation years & $a$ for $\lambda\lambda$=[390-399]nm &
$(\frac{\tau_\kappa}{\tau_S})$ for $\lambda$=399 nm &
$\frac{A_g^F}{A_g^{ghost}}$ \\
 \hline
 1981 & 3.00 & 0.028 & 1.065 \\
 1993 & 1.97 & 0.02 & 1.165 \\
 1995 & 1.85 & 0.019 & 1.188 \\ \hline
 \end{tabular}
\end{center}
\end{table}

In Table 1 we adduce the ratio of the geometric albedo values for the Fraunhofer line KCaII $A_g$ and for its ghost $A_g^{ghost}$ in various years. As the Raman scattering effects decrease when the values of $a$ increase, we can conclude that the observational variations of $\frac{A_g^F}{A_g^{ghost}}$ are due to
the variation of the aerosol component of the planetary atmosphere.

As we analyzed $a$ variations in various years and the latitude of the center of the visible disk variations, it is readily seen that $a$ decreases when the latitude of the visible disk center decreases (from -65$^o$ in 1981 to -35$^o$ in 1995). This regularity is confirmed by Dementiev \cite{Dementiev1, Dementiev2}
and Rages et al \cite{Rages}. They claim that the aerosol abundance was much less in Uranus' atmosphere in 1961-1973 than in 1981-1983. The latitude of the visible disk center in 1961 was -22$^o$ while in 1981 it was -65$^o$. The qualitative explanation of this effect was given by different authors \cite{Karkoschka3, Smith, Sromovsky2}. They have observed that the aerosol cloud features mostly occupied the moderate latitudes.

\section*{Conclusions}
\indent \indent In this paper we determined the spectral values of $(\frac{\tau_a}{\tau_R})$ and $(\frac{\tau_\kappa}{\tau_S})$ for Uranus, using the method of the optical parameter determinations with the use of the Raman scattering details for nonisothermal atmosphere \cite{Kostogryz, Morozhenko4}.

Using this data we can make the following conclusions:
\begin{enumerate}
 \item the values of $(\frac{\tau_a}{\tau_R})$ and
$(\frac{\tau_\kappa}{\tau_S})$ vary in time and depend on the
horizontal inhomogeneity of the Uranus' disk;
\item the possible cause of the long-period variations of the
Uranus' geometrical albedo spectra, which also depend on the latitude,
 is inhomogeneous aerosol distribution.
\end{enumerate}

\section*{Acknowledgements}
\indent \indent I thank A.V.Morozhenko and A.P.Vidmachenko for good advices and
useful discussion of obtained results. And also I express my appreciation to Kathy Rages for a very cogent review and useful discussion.


\begin{thebibliography}{20}
{\small
\bibitem{Cochran}Cochran W.D., Trafton L.M. Astrophys.J., V. 219, 1, pp.756-762 (1978)
\bibitem{Courtin}Courtin R. Planetary and Space Science, V. 47, 8-9, pp.1077-1100 (1999)
\bibitem{Delbouille}DelbouilleL. Roland G., Neven L., Liege:Univ.press (1973)
\bibitem{Dementiev1}Dementiev M. Kinematika i physika nebesnyh tel, V. 8, 2, pp.25-35 (1992)
\bibitem{Dementiev2}Dementiev M. PhD Thesis, pp.220 (1999)
\bibitem{Karkoschka1}Karkoschka E. Icarus, V. 111, pp.967-982 (1994)
\bibitem{Karkoschka2}Karkoschka E. Icarus, V. 133, pp.134-146 (1998)
\bibitem{Karkoschka3}Karkoschka E. Science, V. 280, 5363, pp.570-572 (1998)
\bibitem{Kostogryz}Kostogryz N.M. Kinematika i physika nebesnyh tel, V. 22, pp.254-259 (2006)
\bibitem{Lindal}Lindal G.F., Lyons J.R., Sweetnam D.N., et al. J.Geophys.Res., A92, 3, pp.14987-15001
(1987)
\bibitem{Morozhenko1}Morozhenko A.V. Kinematika i physika nebesnyh tel, V. 13, 4, pp.20-29 (1997)
\bibitem{Morozhenko2}Morozhenko O.V. Naukova Dumka, p.206 (2004)
\bibitem{Morozhenko3}Morozhenko A.V. Kinematika i physika nebesnyh tel, V. 22, 2, pp.138-153 (2006)
\bibitem{Morozhenko4}Morozhenko A.V., Kostogryz N.M. Kinematika i physika nebesnyh tel, V. 21, 2, pp.114-120 (2005)
\bibitem{Neff}Neff J.S. et al. V. 60, pp.221-235 (1984)
\bibitem{Pollack}Pollack J.B., Rages K., Baines K.H., et al. Icarus, V. 65, 2/3, pp.442-466 (1986)
\bibitem{Rages}Rages K.A., Pollack J.B., Tomasko M.G., Doose L.R. Icarus, V. 89, pp.359-376 (1991)
\bibitem{Smith}Smith B.A. et al. Science, V. 233, 4759, pp.43-64 (1986)
\bibitem{Sromovsky1}Sromovsky L.A. V. 173, pp.254-283 (2005)
\bibitem{Sromovsky2}Sromovsky L.A., Fry P.M. Icarus, 179, pp.459-484 (2005)
}
\end{thebibliography}
\end{document}